\documentclass[11pt]{article}

\usepackage{graphicx}
\usepackage{amsmath,amsfonts,amssymb,amsthm}

\usepackage{times}
\usepackage{mathptmx}
\usepackage{moreverb}

\advance\evensidemargin by -1.1cm
\advance\oddsidemargin by -1.1cm
\advance\textwidth by 2.2cm

\advance\topmargin by 1.1cm
\advance\textheight by 2.2cm

\begin{document}

\title{An Experimental Analysis of
Lemke-Howson Algorithm}
\author{
Bruno Codenotti\thanks{IIT-CNR, Via Moruzzi 1, 56124 Pisa, Italy. Email:
bruno.codenotti@iit.cnr.it.} \and Stefano De Rossi\thanks{SSSA,
P.zza Martiri della Libert\`a 33, 56127 Pisa, Italy. Email:
s.derossi@sssup.it.} \and Marino
Pagan\thanks{SSSA, P.zza Martiri della Libert\`a 33, 56127 Pisa,
Italy. Email: m.pagan@sssup.it.}}
\date{}
\maketitle

\begin{abstract}
We present an experimental investigation of the performance of the
Lemke-Howson algorithm, which is the most widely used algorithm for
the computation of a Nash equilibrium for bimatrix games.
Lemke-Howson algorithm is based upon a simple pivoting strategy,
which corresponds to following a path whose endpoint is a Nash
equilibrium. We analyze both the basic Lemke-Howson algorithm and a
heuristic modification of it, which we designed to cope with the
effects of a `bad' initial  choice of the pivot. Our experimental
findings show that, on uniformly random games, the heuristics
achieves a linear running time, while the basic Lemke-Howson
algorithm runs in time roughly proportional to a polynomial of
degree seven. To conduct the experiments, we have developed our own
implementation of Lemke-Howson algorithm, which turns out to be
significantly faster than state-of-the-art software. This allowed us
to run the algorithm on a much larger set of data, and on instances
of much larger size, compared with previous work.
\end{abstract}

\section{Introduction}

The computation of a Nash equilibrium for bimatrix games has
attracted a lot of attention in recent years. The problem is of
central importance in several theoretical and applied areas, and has
many applications in different fields, like the social sciences,
biology, economics, etc.

The computational complexity of this problem has been unknown for
many years, to the point that in 2001 Papadimitriou \cite{pap}
mentioned it as one of the most important open problems in
computational complexity. Recently the problem has been proved
complete for a complexity class, $PPAD$, which contains problems for
which efficient algorithms are not believed to exist
\cite{chendeng}.

In spite of the interest on the problem, little work has been done
to carry out an accurate evaluation of the performance of the
algorithms actually used to solve it. In this paper we
experimentally analyze Lemke-Howson algorithm, which is the best
known algorithm for the computation of a Nash equilibrium for
bimatrix games.

We  provide a new implementation of this algorithm, which turns out
to be significantly faster than state-of-the-art software, and give
an account of its performance. This new implementation allowed us to
experimentally analyze Lemke-Howson algorithm on a much larger set
of sample data, and on instances of much larger size, compared with
previous work.

\begin{center}
\begin{figure}[h]
\begin{minipage}[t]{6cm}
\centering
\includegraphics[width=6cm]{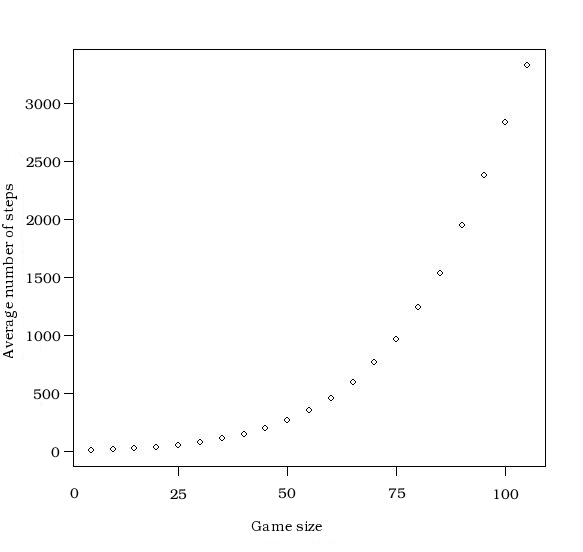}
\caption{Average number of pivoting steps performed by LH as a
function of the game size.}
\label{mean_size}
\end{minipage}
\ \hspace{2mm} \hspace{3mm} \
\begin{minipage}[t]{6cm}
\centering
\includegraphics[width=6cm]{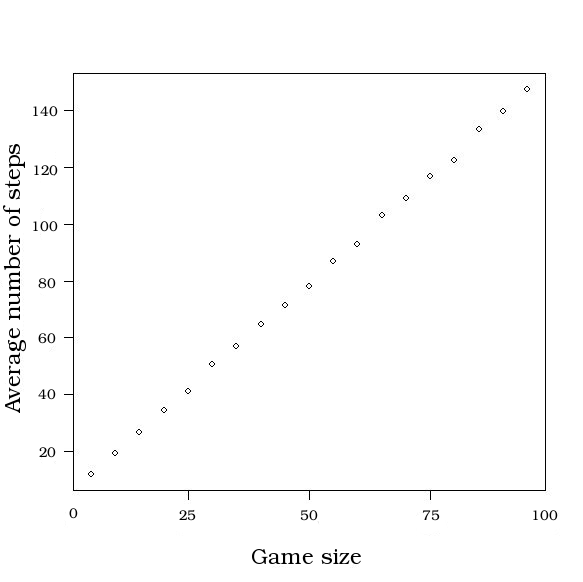}
\caption{Average number of pivoting steps performed by our heuristics
as a function of the game size.}
\label{mean_size_heu}
\end{minipage}
\end{figure}
\end{center}

We also develop a heuristic modification of Lemke-Howson
algorithm, which reduces the computational inefficiency which may
result from a `bad' initial choice of the pivot. On uniformly random
games, this heuristics significantly outperforms the basic
Lemke-Howson algorithm. While Lemke-Howson takes a number of steps which is roughly
proportional to a
polynomial of degree 7 with respect to the game size (see Figure
\ref{mean_size}), the heuristics takes a linear number of steps
(see Figure \ref{mean_size_heu}).

This improvement makes it possible to compute equilibria in a reasonable time 
in games with size well beyond 1000 $\times$ 1000. The details can be found in the Appendix.
Note that with standard Lemke-Howson algorithm, one can compute equilibria only on games
of size up to a few hundreds (see Figure \ref{noi_gambit}).

\subsection{Other algorithms and previous experimental results}

Lemke-Howson algorithm (LH from now on) is the most widely used
algorithm for the computation of Nash equilibria in bimatrix games
\cite{lh}. Nice descriptions of LH can be found in
\cite{shapley,vstengel}.

More recently, new algorithms have been developed. Porter, Nudelman
and Shoham \cite{simple} introduced a simple search method
(\emph{PNS}) based on the enumeration of all strategy supports.
Sandholm, Gilpin and Conitzer introduced a different algorithm
(\emph{MIP}), based on \emph{Mixed Integer Programming}
 \cite{mip}.

A taxonomy of games can
be found in \cite{gamut}. The results obtained in \cite{gamut,simple}
show that two of the
most challenging classes of games are the ``uniformly random games''
and the ``covariant games'', which are the main objects
of investigation in our paper.

Previous experimental works \cite{simple,mip} have shown that on
games with small and balanced support, such as random games,
\emph{PNS} outperforms both \emph{MIP} and LH, while on games with
medium-size support LH outperforms both \emph{MIP} and \emph{PNS}.
These experimental findings have been obtained by using the
implementation of LH released in Gambit. A major difference with our
work is that we have performed the experiments on a much larger set
of sample data, and on instances of much larger size (see
Section~\ref{expres}).

\subsection{Organization of this paper}

In Section 2 we give some background, and introduce the
notation used in the paper. In Section 3 we briefly describe our
implementation, giving account of its performance, and present our
experimental results. In Section 4 we describe our
heuristics, and give ample evidence of its efficiency on uniformly
random games. In the Appendix we give a detailed description of our
implementation of LH, and present some further experimental results
on some classes of games, which, for lack of space, we had to omit
from this extended abstract. The source code of our implementation 
can be found at http://allievi.sssup.it/game.

\section{Background and Notation}

We consider bimatrix games in {\em normal form}. These games are
described in terms of two matrices, containing the {\em payoffs} of
the two players. The rows (resp. columns) of both matrices are
indexed by the row (resp. column) player's {\em pure strategies}.

A {\em mixed strategy} consists of a set of pure strategies and a
probability distribution (a collection of nonnegative weights adding
up to one) which indicates how likely it is that each pure strategy
is played. In other words, each player associates to her $i$-th pure
strategy a number $p_i$ between $0$ and $1$, such that $\sum_i p_i
=1$.

The pure strategies played with positive probability form the {\em
support} of a mixed strategy.

Let us consider a two-player game, where the row (resp., column)
player has $m$ (resp., $n$) pure strategies, and let $x$ be a mixed
strategy of the row player, and $y$ a mixed strategy of the column
player. Strategy $x$ is the $m$-tuple $x = (x_1, x_2, \dots, x_m)$,
where $x_i \geq 0$, and $\sum_{i=1}^m x_i = 1$. Similarly, $y =
(y_1, y_2, \dots, y_n)$, where $y_i \geq 0$, and $\sum_{i=1}^n y_i =
1$. Let now $A=(a_{ij})$ be the payoff matrix of the row player. The
entry $a_{ij}$ is the payoff to the row player, when she plays her
$i$-th pure strategy and the opponent plays the pure strategy $j$.
According to the mixed strategies $x$ and $y$, the entry $a_{ij}$
contributes to the expected payoff of the row player with weight
$x_iy_j$. The expected payoff of the row player can be evaluated by
adding up all the entries of $A$ weighted by the corresponding
entries of $x$ and $y$, i.e., the payoff is $\sum_{ij} x_iy_j
a_{ij}$. This can be rewritten as
 $\sum_{i} x_i \sum_j a_{ij} y_j$, which can be expressed in matrix
terms as\footnote{We
 use the notation $x^T$ to denote the transpose of
vector $x$.}$x^TAy$. Similarly, the expected payoff of the column
player is $x^TBy$.

A pair $(x, y)$ is a {\em Nash equilibrium} if $x^TAy \geq
{x'}^TAy$, and $x^TBy \geq x^TBy'$, for all stochastic vectors $x'$
and $y'$. If the pair $(x, y)$ is a Nash equilibrium, we say that
$x$ (resp. $y$) is a {\em Nash equilibrium strategy} for the row
(resp. column) player. It is well known that a Nash equilibrium in
the mixed strategies always exists \cite{nash}.

An equivalent definition is the following. A Nash equilibrium for a
bimatrix game $(A,B)$ is a pair of {\em mixed strategies} $(x,y)$
such that each pure strategy is played with positive probability
only if it is a {\em best response} to the other player's mixed
strategy (linear complementarity constraints):

\begin{equation*}
\left\{
\begin{array}{cc}
\forall i \in \{1,\dots,m\} \text{ either } x_i=0 \text{ or } (A y)_i
\ge max_k (A y)_k ;\\
\forall j \in \{1,\dots,n\} \text{ either } y_j=0 \text{ or } (x^T
B)_j \ge max_k (x^T B)_k .\\
\end{array}
\right.
\end{equation*}

To the set of equilibria, we add the {\em artificial equilibrium}
$(0,0)$, where no strategy is played. It satisfies the linear
complementarity constraint, but it is not a valid equilibrium.

We also say that a pair of mixed strategies form a {\em
quasi-equilibrium} if all but one of the pure strategies played with
positive probability satisfy the linear complementarity constraint.

Given a bimatrix game, and starting from any equilibrium (including
the artificial equilibrium), LH follows a path in a graph whose
vertices are the equilibria and the quasi-equilibria. Thus, starting
from the artificial equilibrium, it is possible to reach a Nash
equilibrium of any given game (see \cite{vstengel} for a detailed
description of this process).

At each step, the state of the algorithm can be represented by the
following system of inequalities, which define the space of the
feasible mixed strategy profiles:

\begin{equation*}
\left\{
\begin{array}{cccc}
\forall i \in \{1,\dots,m\} \text{ } x_i \ge 0 ;\\
\forall j \in \{1,\dots,n\} \text{ } y_j \ge 0 ;\\
\forall j \in \{1,\dots,n\} \text{ } (x^T B)_j \le 1 ;\\
\forall i \in \{1,\dots,m\} \text{ } (A y)_i \le 1 ;\\
\end{array}
\right.
\end{equation*}

Note that the formulation above has been obtained by applying a
suitable scaling procedure where the maximum payoff is set to $1$,
and the mixed strategies are scaled accordingly.

Following \cite{vstengel}, these inequalities can be rewritten as:

\begin{equation*}
\begin{pmatrix}

a_{11} & \dots & a_{1n}\\
\vdots & \ddots & \vdots\\
a_{m1} & \dots & a_{mn}\\
-1 & \dots & 0\\
\vdots & \ddots & \vdots\\
0 & \dots & -1\\

\end{pmatrix}
\begin{pmatrix}

y_1\\
\vdots\\
y_n\\

\end{pmatrix}
\le
\begin{pmatrix}

1\\
\vdots\\
1\\
0\\
\vdots\\
0\\

\end{pmatrix}
\end{equation*}

\begin{equation*}
\begin{pmatrix}

x_1 & \dots & x_m

\end{pmatrix}
\begin{pmatrix}

-1 & \dots & 0 & b_{11} & \dots & b_{1n}\\
\vdots & \ddots & \vdots & \vdots & \ddots & \vdots\\
0 & \dots & -1 & b_{m1} & \dots & b_{mn}\\

\end{pmatrix}
\le
\begin{pmatrix}

0 & \dots & 0 & 1 & \dots & 1\\

\end{pmatrix}
\end{equation*}

In order to describe the state of the algorithm at each step, we
will use a structure called {\em tableau}, which consists of a
representation of the inequalities above as equalities, obtained by
introducing {\em slack variables}.

The transformation of inequalities into equalities of the tableau
occurs as follows. Consider for example the inequality
\begin{equation*}
a_{11} y_1 + a_{12} y_2 + \dots + a_{1n} y_n \le 1
\end{equation*}

After introducing the slack variables, this inequality becomes:
\begin{equation*}
a_{11} y_1 + a_{12} y_2 + \dots + a_{1n} y_n + r_1 = 1
\end{equation*}

Finally the equation is represented with all the non basic variables
on the right hand side. Therefore, at the first step, we will have:
\begin{equation*}
r_1 = 1 - a_{11} y_1 - a_{12} y_2 - \dots - a_{1n} y_n.
\end{equation*}

In this equation, the variable on the left hand side is the basic
variable, and the constant on the right hand side represents its
current value. Starting from this representation, the algorithm
proceeds via {\em complementary pivoting} steps, which allow the
computation to move from a quasi-equilibrium to another one by
changing each time one of the basic variables. This translates into
a corresponding change in the tableau. The choice of the new basic
variable is determined by the {\em minimum ratio test} - lines
12-15 of the pseudocode, which is in the Appendix.
The complementary pivoting procedure terminates
when all the strategies satisfy the complementary constraint, i.e.,
when an equilibrium is reached.

\section{Experimental Results}
\label{expres}

\subsection{Our implementation}
The current state-of-the-art software for the computation of
equilibria for games is Gambit \cite{gamb}. It can be used to deal
with extensive and normal form games. In particular, it is endowed
with suitable command-line tools which allow the user to compute
Nash equilibria for games given in normal form.

Being intended as a tool for high-level game theoretical analysis,
Gambit does not provide accurate information on the efficiency of
LH, e.g., on the number of complementary pivoting steps, and on
properties of each execution. Therefore Gambit does not seem to be
suitable for large scale experiments.

The need of low-level control on the computation and of a faster
software tool lead us to develop our own implementation of LH. A
detailed description of our implementation is given in the Appendix.

Figure \ref{noi_gambit} shows the average running time of LH, 
as a function of the game size. The games were
generated using GAMUT\footnote{RandomGame game class.}, a suite of game generators \cite{gamut}.

\begin{figure}[h]
\centering
\includegraphics[width=8.8cm]{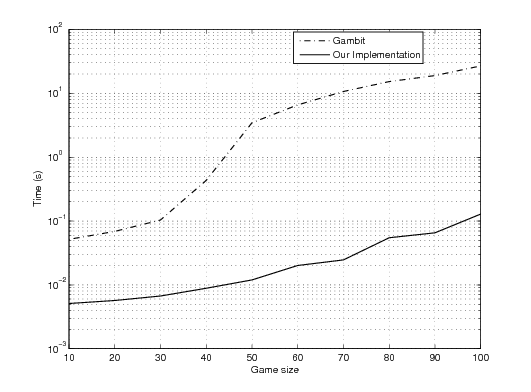}
\caption{Average running time of LH. Our
implementation (solid line) and Gambit (dotted line). On game sizes larger
than 100$\times$100, a relevant fraction of games were not solved by Gambit
within the capping time of 120 secs.}
\label{noi_gambit}
\end{figure}

\subsection{Uniformly Random Games}

In this section we present the data collected for uniformly
random games, i.e., games in which each payoff is chosen at random
from the uniform distribution.
We focus on the number of complementary pivoting steps,
rather than on the execution time, the latter being too much
implementation dependent.

When collecting the data, for each equilibrium computation, we kept
track both of the equilibrium itself (the strategies played and the
probability of playing each of them), and of the number of pivoting
steps needed to reach it.

Figure \ref{mean_size} shows the average number of pivoting steps
performed by LH: one can see that the number of steps grows
polynomially (approximately according to a polynomial of degree 7)
with the size of the game. The data has been obtained by running LH
on 100,000  instances.

\begin{figure}[h]
\centering
\includegraphics[width=6cm]{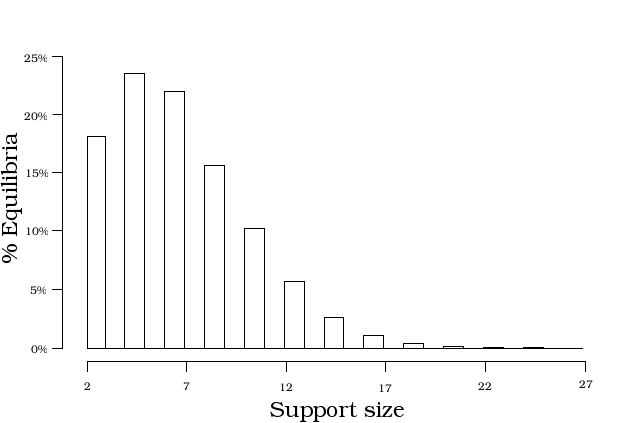}
\includegraphics[width=6cm]{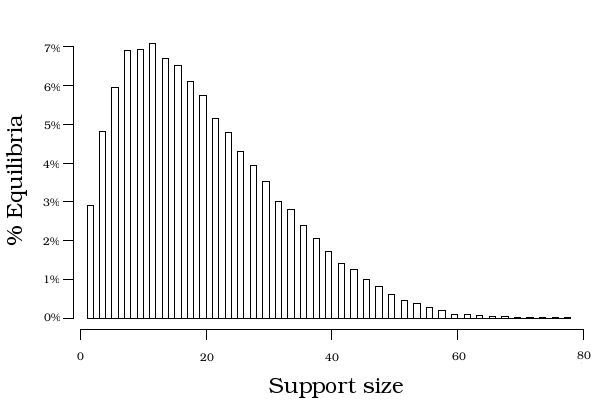}
\caption{Support size distribution for games with 20 strategies per
player (on the left), and 100 strategies per player (on the right). 
We take as support size the sum of the sizes of the supports of the 
strategies of both players.}
\label{supp_dist}
\end{figure}

Figure \ref{supp_dist} shows the distribution of the support size of
equilibria found by LH. We see that there is a small number of
equilibria with large support size. Thus the behavior of LH agrees
with the fact that the probability to find an equilibrium with a
large support size is very low for (sufficiently large) uniformly
random games. Although these findings only concern those equilibria
which are found by LH, they show a close agreement with the
theoretical results on the support of equilibria of random games
\cite{vempala}.

\begin{figure}[h]
\centering
\includegraphics[width=8cm]{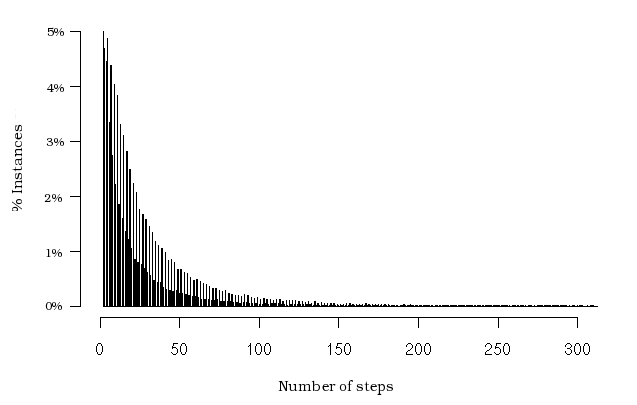}
\begin{tabular}{|cccccc|}
\hline
Mode
& Mean
& 1st Quartile
& 3rd Quartile
& 95\% Quantile
& 99.5\% Quantile \\ \hline
2 & 27.39 & 7 & 34 & 91 & 203\\
\hline
\end{tabular}
\caption{Number of steps performed by LH on uniformly random games
of size 20}
\label{20all}
\end{figure}

To gather some further insights on the behavior of the algorithm, we
analyzed the distribution of the number of pivoting steps performed
by LH  on games of three different sizes: 20, 40, and 100 strategies
per player. For each size, we analyzed a large number of runs. More
precisely, about 7.5 million runs for games with 20 and 40
strategies per player, and 700,000 runs for games with 100
strategies per player. Figure \ref{20all} shows the distribution of the number
of steps for runs on 7.5 million games with 20 strategies per
player.

Figure \ref{20all} shows that the {\em mode} of the distribution of
the number of complementary pivoting steps is $2$, which is the
minimum number of steps needed to reach a pure equilibria. The data
in Figure \ref{20all} illustrates that the distribution is very
sparse: the third quartile is at 34 steps. Thus it is hard to make
accurate predictions on the number of steps based on the game size.
Indeed, in order to gather 99.5\% of our statistical data, we had to
go up to more than $200$ steps for games with 20 strategies per
player. A closer look at the distribution gives us additional
information on
the behavior of the algorithm: two different distributions are
interleaved, one for an even number of steps, and another one for an
odd number of steps. Except for two, three, and four steps, the two
interleaved distributions are almost geometrical.

\begin{figure}[h]
\centering
\includegraphics[width=5.5cm]{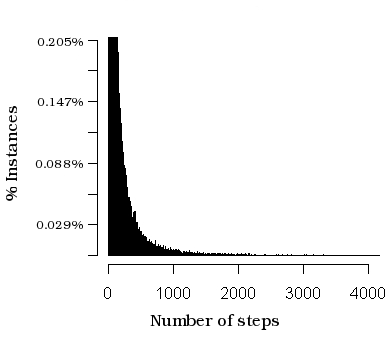}
\includegraphics[width=5.5cm]{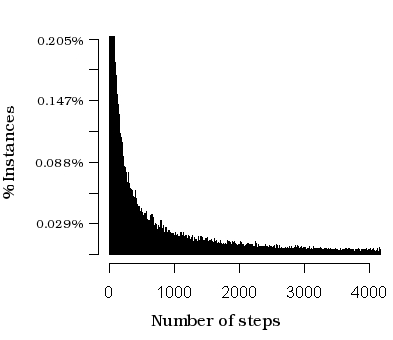}
\caption{Distribution of the number of steps performed by LH on
games with 40 (left) and 100 (right) strategies per player. The y-axis is
intentionally limited to ease the comparison of the two
distributions.} \label{step_dist_2}
\end{figure}

Similar observations hold for larger games. Figure \ref{step_dist_2}
shows the statistical data for games of size $40$ and $100$. Zooming
on the distribution, we see that the two interleaved distributions
still exist. As the games size increases these distributions do not
change qualitatively. Indeed, while the mean and the sparseness of
the distribution increase with the size, the shape stays the same.

\subsection{Other classes of games}

Further experimental results have been obtained on covariant games,
as defined in \cite{covariance}. This class contains families of
instances which are \emph{harder} to solve by LH than uniformly
random games. For lack of space, the details are in the Appendix.

\section{A heuristic improvement on Lemke-Howson algorithm}

The number of steps taken by LH depends on the pivot it initially
chooses. Since this choice is arbitrary, the algorithm might take a
large number of steps, even on instances where there exist pivots on
which it would terminate quickly. The performance of LH could be
significantly improved if one knew in advance which pivot
leads to the minimum number of steps.

In this section, we first analyze the performance of a clairvoyant
algorithm (ND-Lemke-Howson) that chooses the \emph{best} pivot and
then executes LH starting from it. We then review a heuristics
proposed by Porter, Nudelman and Shoham \cite{simple}, which can be
viewed as a simulation of ND-Lemke-Howson, and finally
introduce a novel heuristics, which achieves a more efficient
simulation of ND-Lemke-Howson.

\subsection{ND-Lemke-Howson}

We simulated ND-Lemke-Howson by executing
the standard Lemke-Howson algorithm on all the possible pivots, 
and choosing the path with the minimum
number of steps. The following is the pseudo-code description of
this algorithm.

\noindent
\makebox[\textwidth]{\hrulefill}
\begin{verbatim}
i = Best_pivot_to_start_from();

Lemke-Howson(pivot=i);

return equilibrium;
\end{verbatim}
\hrule\hfill\\

\begin{figure}[h]
\begin{minipage}[t]{6.1cm}
\centering
\includegraphics[width=6cm]{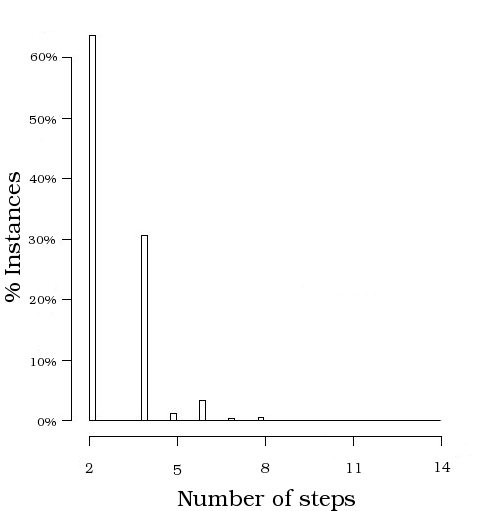}
\caption{Number of steps performed by ND-Lemke-Howson algorithm for
40$\times$40 games.} \label{ndlh}
\end{minipage}
\ \hspace{3mm} \hspace{4mm}\
\begin{minipage}[t]{6.1cm}
\centering
\includegraphics[width=6cm]{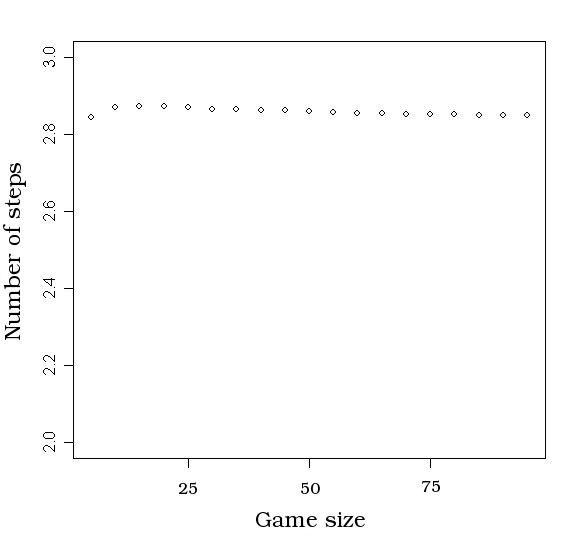}
\caption{Mean of the number of steps performed by
ND-Lemke-Howson as a function of the game size.}
\label{mean_size_nd}
\end{minipage}
\end{figure}

The first predictable result is that the distribution of the number
of steps for ND-Lemke-Howson is much less sparse than the one for
LH: for a large fraction of games it terminates in a very small
number of steps. Figure \ref{ndlh} shows the distribution of the
number of steps after 100,000 executions of ND-Lemke-Howson on
40$\times$40 random games. In particular, notice that the maximum
number of steps is 12.

A significant difference between LH and ND-Lemke-Howson is that the
average number of steps taken by ND-Lemke-Howson does not increase
with the size of the game. The mean is almost stable at 2.8, and it
slightly decreases as the game size increases (compare Figure
\ref{mean_size_nd} with Figure \ref{mean_size}). This fact can be
explained by observing that there is little correlation between the
executions of LH for the same game on different pivots. And so,
since the number of pivots increase linearly with size, the
probability to find a short path increases accordingly.

\subsection{Porter, Nudelman and Shoham heuristic approach}

Porter, Nudelman and Shoham heuristics \cite{simple} simulates
ND-Lemke-Howson by keeping track of all possible different
executions of LH (two times the game size), and then performing a
single pivoting step on each execution, until one of the paths
reaches an equilibrium; the overhead with respect to ND-Lemke-Howson
is approximately given by the number of steps performed by
ND-Lemke-Howson multiplied by the number of possible pivots. The
following is the pseudo-code implementation of this heuristics.

\noindent
\makebox[\textwidth]{\hrulefill}
\begin{verbatim}
create 2 * dim different tableaux

while( true )
{
  for i = 1 to 2 * dim do
  {
    pivoting_step( tableaux[i] );
    if an equilibrium is found then
       return equilibrium;
    else
       continue;
  }
}
\end{verbatim}
\hrule
\hfill\\

A significant implementation issue with this heuristics is that it
requires a large amount of memory. Indeed, we have to store a
different tableau for each pivoting path. Therefore the memory
consumption is worse than LH by a factor of $2 \cdot dim$. For a
$dim \cdot dim$ game, one has to store $2 \cdot dim$ tableaux, each
of size quadratic in the game size, so that the overall memory
consumption is cubic in the game size. This heuristics is thus
unsuitable to compute equilibria for games with more
than a few hundreds strategies per player.

\subsection{A novel heuristics}

We developed a slightly different heuristics, which avoids the
problem of a large memory consumption, takes (on average) less
pivoting steps, and is much easier to implement.
This heuristics takes a parameter, which we call \textit{capping},
that tells how
many steps will be performed on each possible pivot before
truncating the execution and starting it again on the following
pivot, until some path reaches an equilibrium. If the length of the paths on every
pivot is larger than the capping value, then we just execute LH on
the last pivot. The following is the pseudo-code implementation
of our heuristics.

\noindent
\makebox[\textwidth]{\hrulefill}
\begin{verbatim}
for i = 1 to 2*dim - 1 do
{
  Lemke-Howson(pivot = i, max_steps = capping);

  if an equilibrium is found then
     return equilibrium;
  else
     continue;
}

if all paths have been truncated then
   Lemke-Howson(pivot = 2*dim, max_steps = INFTY);
   return equilibrium;
\end{verbatim}
\hrule\hfill\\

The memory consumption is clearly the same as in standard LH, since at each
time step we use just one tableau. This allows our implementation to
be executed even on games with thousands of strategies for each player.
What makes this heuristics efficient is the fact that only a small fraction
of games does not admit any short path.

\begin{figure}[h]
\begin{minipage}[t]{6.1cm}
\centering
\includegraphics[width=5.6cm]{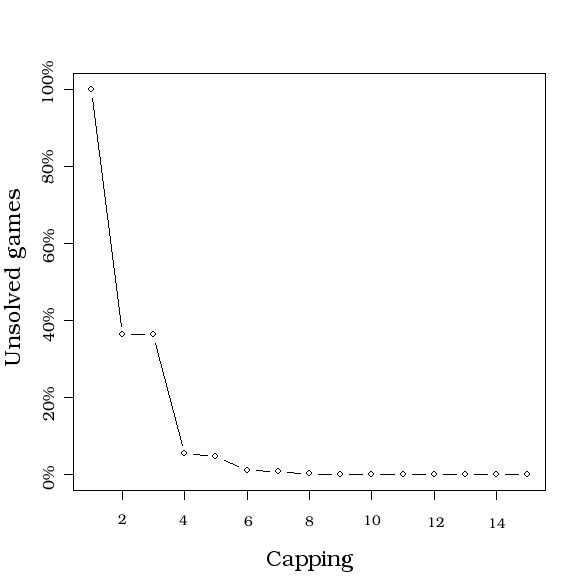}
\caption{Fraction of 100$\times$100 games for which the heuristics is
forced to execute LH on the last pivot, with respect to the
\emph{capping} parameter.}
\label{perc_capping}
\end{minipage}
\ \hspace{3mm} \hspace{4mm} \
\begin{minipage}[t]{6.1cm}
\centering
\includegraphics[width=5.6cm]{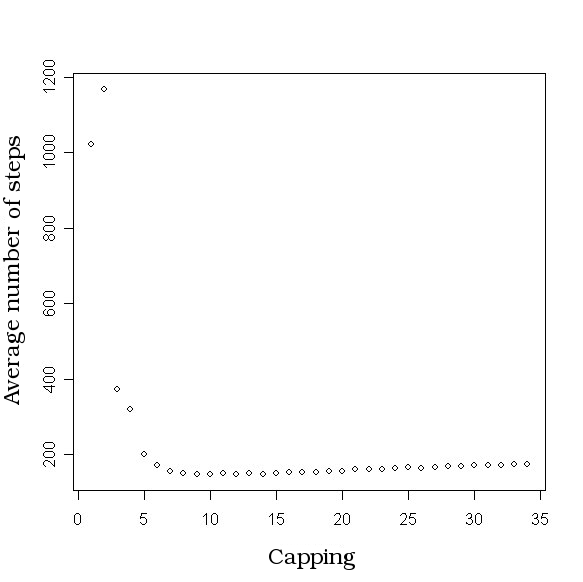}
\caption{Average number of steps performed by our heuristics as a
function of the capping parameter on 100$\times$100 games.}
\label{steps_cap}
\end{minipage}
\end{figure}

Figure \ref{perc_capping} shows the fraction of games (of size
100$\times$100) for which the algorithm does not terminate before
reaching the last pivot, as a function of the capping value. The lower
is the capping value, the higher will be the probability to pivot on
the last strategy. We can see that by choosing a capping value greater
 than 20 this probability is very small.

We performed an extended analysis to determine the best capping
value for each game size on uniformly random games, and we have seen
that the best value is approximately of 10 steps, and does not
significantly change increasing the game size. Figure
\ref{steps_cap} shows how the average number of steps varies with
the capping value for 100$\times$100 games. The performance improves
significantly as the capping value approaches 10, and then gets
slowly worse for larger values. The best performance (on average) is
obtained by choosing a capping value of 10.

\begin{figure}[h]
\centering
\includegraphics[width=6cm]{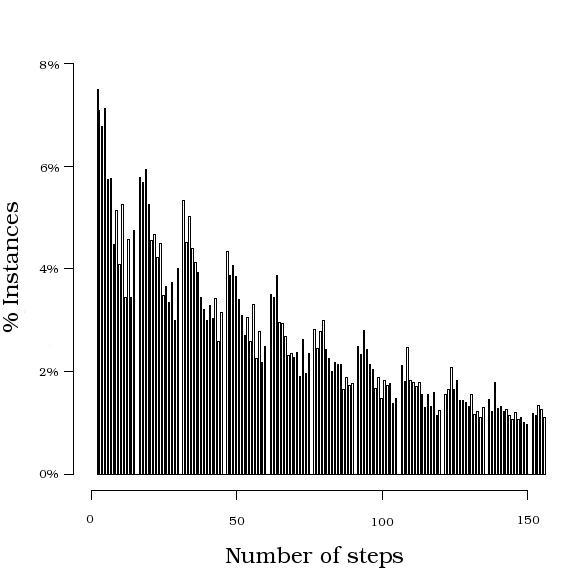}
\caption{Number of steps performed by our heuristics, on
100$\times$100 games.}
\label{100all_heu}
\end{figure}

Figure \ref{100all_heu} shows the distribution of the number of
steps for our heuristics. It is similar to the distribution obtained
for LH, although it is less sparse and not monotonic. Inside of it
one can recognize many geometrical distributions, which represent
the distribution of the feasible short paths that LH can generate
from each pivot.

The data in Figure \ref{mean_size_heu} shows the average number of
steps of our algorithm. The average number of steps turns out to be
linear, which is a drastic reduction compared with what we observed
for LH (Figure \ref{mean_size}).

\newpage

\appendix

\section{Our implementation}

Our implementation of LH can either search for a Nash equilibrium in a normal
form bimatrix game, executing LH on a given pivot strategy, or it
can enumerate all equilibria reachable by LH starting from the
artificial equilibrium.

We will only describe the first feature, the latter being of less
importance to the experimental analysis we carry out in this paper.

\subsection{The algorithm}

\subsubsection{Data Structures}

The most relevant data structures are those used to store tableaux,
equilibria, and lists of equilibria.

Since our main concern is execution time rather than memory
consumption, we do not use a sparse matrix implementation of the
tableau, but just a naive matrix representation. This allows us to
efficiently access and update the tableau. To keep things simple,
instead of using an array to keep track of the strategies which are
in the basis, we store this information directly in the first column
of the tableau. The second column represents the actual value of the
variable in the basis for that row, while all the other entries
represent the coefficients of all nonbasic variables.

For the sake of the reader, we now show an example of how a tableau
looks like after being initialized. Consider the following game:

\begin{equation*}\label{bim}
  A = \left[
    \begin{array}{ c c }
      1 & 2 \\
      3 & 4 \\
      5 & 6
    \end{array}
    \right]
  \hskip1cm B = \left[
    \begin{array}{c c}
      7 & 8\\
      9 & 10\\
      11 & 12
    \end{array}
    \right]
\end{equation*}

The linear complementarity formulation is the following:

\begin{equation*}
\left\{
  \begin{array}{llllll}
    s_1 = 1 & { } & { } & { } & -x_4 & -2x_5\\
    s_2 = 1 & { } & { } & { } & -3x_4 & -4x_5\\
    s_3 = 1 & { } & { } & { } & -5x_4 & -6x_5\\
    s_4 = 1 & -7x_1 & -9x_2 & -11x_3 & { } & { }\\
    s_5 = 1 & -8x_1 & -10x_2 & -12x_3 & { } & { }
  \end{array}
\right.
\end{equation*}

\noindent where $s_1,\ldots, s_5$ are the slack variables, and $x_1,
\ldots, x_5$ are the actual variables. This is the initial set up,
where the equations represent the artificial equilibrium, with all
the slack variables in the basis. The tableaux will be:

\begin{equation*}
Tableau_1 = \left[
  \begin{array}{ccccccc}
    -1 & 1 & 0 & 0 & 0 & -1 & -2\\
    -2 & 1 & 0 & 0 & 0 & -3 & -4\\
    -3 & 1 & 0 & 0 & 0 & -5 & -6
  \end{array}
\right]
\end{equation*}
\begin{equation*}
Tableau_2 = \left[
  \begin{array}{ccccccc}
    -4 & 1 & 0 & 0 & -7 & -9 & -11\\
    -5 & 1 & 0 & 0 & -8 & -10 & -12
  \end{array}
\right]
\end{equation*}

We broke the tableau in two smaller independent tableaux, to
simplify its update during the complementary pivoting steps. The
first column represents the {\it index} of the basic variable for
the given row. Positive indices are used for actual variables and
negative ones for slack variables. The second column stores the
value of that variable. All the other entries in the tableau are the
coefficients of all the other variables, which are out of the basis.
First come the coefficients of the slack variables, then those of
the actual variables. For example, in the first tableau
(representing the first three linear complementarity equations)
columns 3 to 5 represent the coefficient of slack variables of
indices 1 to 3 (these are all zeros, because all slacks are in the
basis in the artificial equilibrium), while columns 6 and 7 are the
coefficients of variables 4 and 5. All the other data structures of
interest are used in the enumeration of all Nash equilibria
reachable by LH. We use lexicographically sorted linked lists to
store both equilibria and lists of equilibria, thus minimizing the
time needed to check if an equilibrium had been already inserted in
the list. Since we wish to keep the time to compare equilibria as
low as possible, we used a null pointer implementation of the
artificial equilibrium.

\subsubsection{Pseudo-Code}

In the following, we describe our implementation of LH.

\begin{listing}[1]{1}
lemke_howson( bimatrix, tableaux, startpivot ) {
  pivot = startpivot

  while( true ) {
    cur_tab = get_cur_tableau( pivot )
    col_i = get_col_i( pivot )

    for i = 1 to cur_tabeau.n_rows {
      if( cur_tab[i][col_i] > 0 )
          continue;

      ratio = -cur_tab[i][1] / cur_tab[i][col_i]
      if( ratio < minimum_ratio ) {
        minimum_ratio = ratio
        row_i = i
      }
    }

    var_out = get_variable( cur_tab[row_i] )
    col_i_out = get_col_i( var_out )

    cur_tab[row_i][col_i] = 0
    cur_tab[row_i][col_i_out] = -1
    for i = 1 to cur_tab[row_i].length
      cur_tab[row_i][i] /= -cur_tab[row_i][col_i]

    for i = 1 to cur_tab.n_rows {
      if( cur_tab[i][col_i] != 0 ) {
         for j = 1 to cur_tab[i].length
      cur_tab[i][j]+=cur_tab[i][col_i]*cur_tab[i][j]

         cur_tab[i][col_i] = 0
      }
    }

    pivot = -var_out
    if(pivot == startpivot or pivot == -startpivot)
       break
  }

  equilibrium = get_equilibrium( tableaux )
  return equilibrium
}
\end{listing}

The algorithm acts on the two matrices describing the game, the
strategy to pivot on at the first step, and the two tableaux. In the
first execution the two tableaux will be built from the artificial
equilibrium, while in any subsequent execution, the algorithm can
start from an arbitrarily chosen equilibrium.

The body of the algorithm consists of an infinite loop (lines 4 to
39 above) where the complementary pivoting steps are performed: the
algorithm exits from the cycle when the variable which is going to
leave the basis is either the starting pivot or its corresponding
slack variable (line 37). Indeed, this means we are at an actual
Nash equilibrium. At the end of the pivoting steps, we only have to
extract the equilibrium from the tableau: this is done by looking at
which strategies are in the basis, and what is the value of the
corresponding variable in the tableau. We can then recover the
actual Nash equilibrium, after normalizing these values so that the
sum of the strategies played by each player is 1. This task is
performed by the get\_equilibrium function, called after the
pivoting steps at line 41.

We now analyze the implementation of the complementary pivoting
steps. We first select the tableau which contains our pivot, and
then determine the column which corresponds to our pivoting strategy
(lines 5 and 6).

The complementary pivoting step consists of two phases: (1)
determining the variable going out of the basis (and the row in the
tableau associated with it), and (2) updating the tableau according
to the new basis. In phase (2), the update involves both the row in
the tableau determined in the phase (1), and the rest of the
coefficients, on the grounds of the expression of our new basic
variable as a linear combination of nonbasic variables.

The minimum ratio test (lines 8 to 17) is done by evaluating the
ratio between the value of all the variables in the basis and the
coefficient of the variable entering the basis (our pivot). This
ratio is obviously calculated only in those rows where the
coefficient of the pivot is negative (lines 9-10). The row which
minimizes this ratio is chosen, along with the variable leaving the
basis (lines 13-16).

The update of the tableau is done as follows. First of all we
determine the index of the column of the variable leaving the basis
(lines 19-20). Then, we update the row involved in the change of the
basic variable: to do this we set the coefficient of the pivot
variable to zero (because now it is a basic variable), the
coefficient of the variable going out of basis to -1 (we are, in
fact, moving that variable from the left hand side of the equation
represented by this row to the right hand side). After dividing all
the other coefficients by the coefficient of the pivot, we get the
final values (lines 22 to 25). Finally, we update all the other rows
in the tableau (lines 27 to 34).

The complementary pivoting rule forces the choice of the next
pivoting variable: it will be the complement of the variable which
just exited the basis (line 36). Therefore the algorithm follows a
path of quasi-equilibria with a duplicate label, i.e., a label for
which both the variable and its slack are in basis, and a missing
label. The process stops when the variable going out of the basis
will be either our initial pivot or its slack (line 37-38).

\subsection{Performance}

\subsubsection{The computational environment}

We executed our program on a SUN X4200 workstation, with 2 AMD
Opteron 2.6 Ghz dual core processors, 4GB of DDR PC3200 RAM, 1MB L2
Cache per core. This workstation runs Debian GNU-Linux Etch amd64 on
a Xen 3.0.2 virtual machine. The program was compiled with gcc 4.0.3
with '-O2 -mtune=x86-64' as compilation options.

\subsubsection{Some Issues with Gambit implementation}
\label{gambit_impl}

For the sake of comparing our implementation of LH with the one
provided within Gambit, we had to make some minor modifications to
Gambit code. In our experiments, we sometimes execute a single LH
run pivoting on a given variable. On the other hand, Gambit's tool
gambit-lcp only allows the user to enumerate all equilibria
reachable by LH starting from the artificial equilibrium. Therefore,
we modified Gambit code, and added the option above. Moreover, we
made other minor changes in order to gather more detailed
information on the execution of the algorithm, i.e. variables
entering and leaving the basis and number of pivoting steps
performed by the algorithm itself.

\subsubsection{Performance of our heuristics}

Figure \ref{tabheu} shows the performance of our heuristics, on uniformly random games. 

\begin{figure}[h]
\centering
\begin{tabular}{|c|c|}
\hline
\textbf{Game Size} & \textbf{Running Time} \\
\hline
200 & 0.1699s\\
300&0.6664s\\
400&1.6184s\\
500&3.1029s\\
600&5.1076s\\
700&8.7794s\\
800&10.4485s\\
900&16.7848s\\
1000&27.7215s\\
1100&32.3306s\\
1200&57.6926s\\
\hline
\end{tabular}
\caption{Average running time of our heuristics with respect to the game size.}
\label{tabheu}
\end{figure}

One can see that, on a standard PC, one can compute equilibria of games with size 
up to 1200$\times$1200 under 1 min. 

\section{Other classes of games}

\subsection{Covariance Games}
\begin{figure}[h]
\centering
\includegraphics[width=7cm]{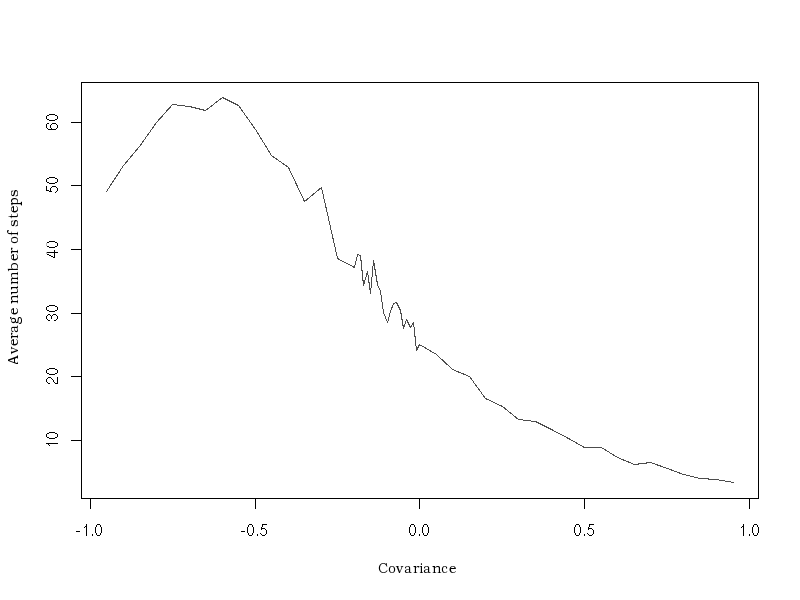}
\caption{Average number of steps performed by LH on a 20$\times$20 covariant
game with respect to the covariance.} \label{covariance20}
\end{figure}

Following \cite{gamut}, we analyzed the performance of LH on the random game model of Rinott
and Scarsini (\cite{covariance}), in which the payoffs are drawn
from two multivariate normal random distribution with a covariance
parameter \emph{$\rho$} varying in the range [-1, 1]. A covariance
of 1 means that the two players share the same payoffs, while a
covariance of -1 means that the game is zero-sum and the payoffs
have a minimal correlation. For a game of size 20, the behavior of
LH changes as shown in Figure \ref{covariance20}.

When $\rho >0$, the number of steps of LH decrease as $\rho$
increases. This might be explained by the theoretical analysis in
\cite{covariance}, which shows that for a positive value of $\rho$,
the probability of finding a pure strategy equilibrium increases as
a monotonic function of $\rho$. When $\rho < 0$, the
mean, mode and median of the distribution of the number of steps
increase dramatically as $\rho$ decreases, reaching a maximum for
$\rho = -0.7$. Similar
results hold for games of different size.

\begin{figure}[h]
\centering
\includegraphics[width=5.25cm]{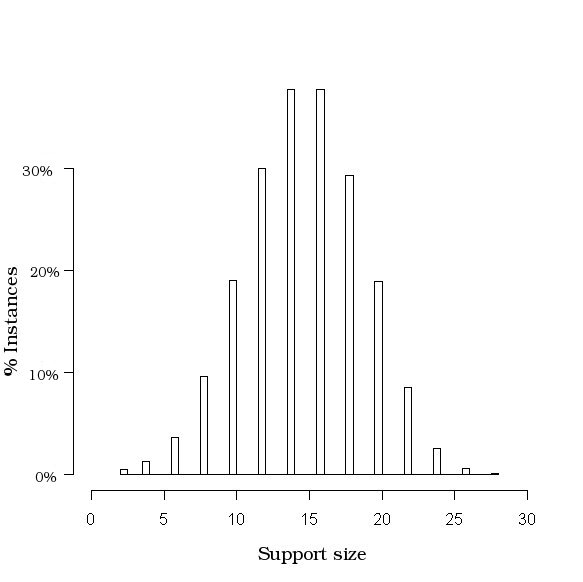}
\includegraphics[width=5.25cm]{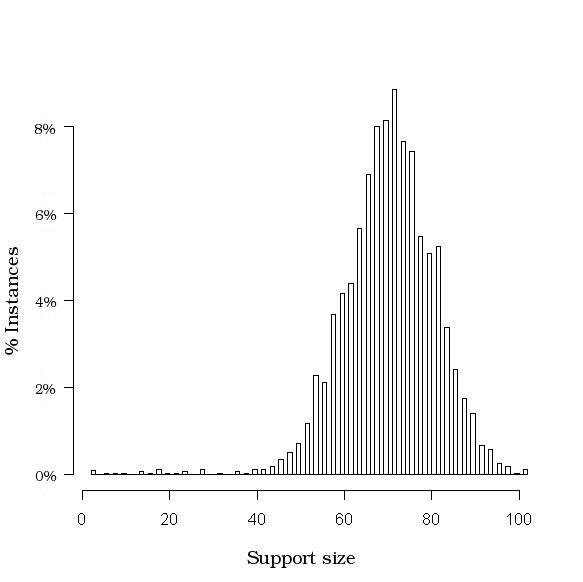}
\caption{Support size distribution for covariant games with 20 strategies per
player (on the left), and 100 strategies per player (on the right) for
$\rho = -0.7$.}
\label{covsup}
\end{figure}

It is interesting to look at the distribution of the support size of
the equilibria reached by LH on 20$\times$20 and 100$\times$100 games
for $\rho =
 -0.7$ (Figure \ref{covsup}). The graph resembles a normal distribution
centered at a value which is slightly less than half the size of the game.

\begin{figure}[h]
\centering
\includegraphics[width=5.25cm]{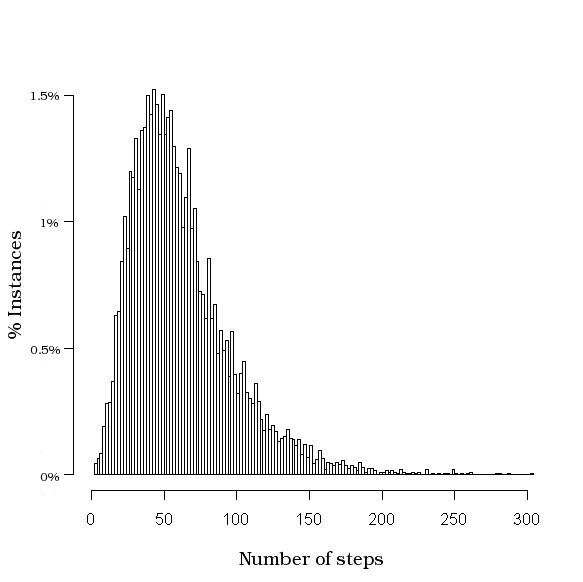}
\includegraphics[width=5.25cm]{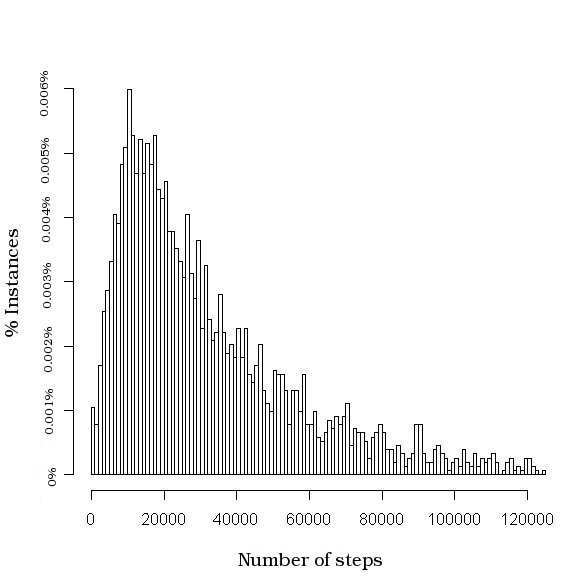}
\caption{Distribution of the number of the steps for covariant games
with 20 strategies per
player (on the left), and 100 strategies per player (on the right) for
$\rho = -0.7$.}
\label{covdis}
\end{figure}

Comparing Figure \ref{covsup} with Figure \ref{supp_dist}, we see that
the equilibria found by LH
on covariant games tend to have a greater average support size.
The resulting distribution of the number of steps turns out to be
shifted to the right
with respect to what happens for uniformly random games (see Figure
\ref{covdis}).

Therefore, it is unlikely that any heuristics of the kind discussed
in this paper could possibly do better than LH.

\end{document}